\documentclass[12pt,preprint]{aastex}

\begin{document}

\title{Gas Giant Protoplanets Formed by Disk Instability in Binary Star Systems}

\author{A. P.~Boss}
\affil{Department of Terrestrial Magnetism, Carnegie Institution of
Washington, 5241 Broad Branch Road, NW, Washington, DC 20015-1305}

\authoremail{E-mail: boss@dtm.ciw.edu}

\vspace{1.0in}

\begin{abstract}

 Roughly half of nearby primary stars are members of binary or multiple
systems, so the question of whether or not they can support the formation of
planetary systems similar to our own is an important one in the search for
life outside our solar system. Previous theoretical work has suggested 
that binary star systems might not be able to permit the formation of gas
giant planets, because of the heating associated with shock fronts driven 
in the stars' protoplanetary disks by tidal forces during the periodic 
close encounters between the two stars. As a result, the disks could become 
too hot for icy bodies to exist, thereby preventing giant planet formation 
by the core accretion mechanism, and too hot for giant planets to form by 
the disk instability mechanism. However, gas giant planets have been 
discovered in orbit around a number of stars that are members of binary 
or triple star systems, with binary semimajor axes ranging from 
$\sim$ 12 to over 1000 AU. We present here a suite of three 
dimensional radiative gravitational hydrodynamics models suggesting 
that binary stars may be quite capable of forming planetary systems 
similar to our own. One difference between the new and previous 
calculations is the inclusion of artificial viscosity in the
previous work, leading to significant conversion of disk kinetic energy into
thermal energy in shock fronts and elsewhere. New models are presented
showing how vigorous artificial viscosity can help to suppress clump 
formation. The new models with binary companions do not employ
any explicit artificial viscosity, and also include the third (vertical)
dimension in the hydrodynamic calculations, allowing for transient phases
of convective cooling. The new calculations of the evolution of initially 
marginally gravitationally stable disks show that the presence of 
a binary star companion may actually help to trigger the formation of
dense clumps that could become giant planets. Earth-like planets 
would form much later in the inner disk regions by the traditional
collisional accumulation of progressively larger, solid bodies. 
We also show that in models without binary companions, which begin
their evolution as gravitationally stable disks, the disks evolve to 
form dense rings, which then break-up into self-gravitating clumps. These 
latter models suggest that the evolution of any self-gravitating 
disk with sufficient mass to form gas giant planets is likely to lead to a 
period of disk instability, even in the absence of a trigger such as a 
binary star companion.

\end{abstract}


\section{Introduction}
 
 Duquennoy \& Mayor (1991) found that only 1/3 of the 164 G dwarf
primary stars within 22 pc of the sun might be true single stars, that is,
stars having no companions more massive than 0.01 $M_\odot$. About 2/3 of 
the G dwarf primaries are thus members of binary or multiple star systems.
The binary frequency for M dwarf primaries within 20 pc is somewhat lower,
no more than 1/2 (Fischer \& Marcy 1992). As a result, it appears that
roughly half the nearby primary stars are single stars. Considering that 
the other half have at least one more stellar companion, less than a third 
of all of the nearby stars are single stars like the sun. Given the drive
to detect and characterize Earth-like planets around the closest stars,
it is clear that binary stars need to be as thoroughly scrutinized as 
single stars to see if they might also be hospitable abodes for 
habitable planets.

 Binary stars have been included on radial velocity planet searches for 
quite some time, beginning with the pioneering search by Walker et al. 
(1995). Over 20 years of data has strengthened the case for a planet with a 
minimum mass of 1.7 $M_J$ (Jupiter masses) orbiting with a semimajor axis
of 2.13 AU around $\gamma$ Cephei A (Walker et al. 1992; Hatzes et al. 2003). 
The $\gamma$ Cephei binary system has an orbital period of $\sim$ 57 yrs,
implying an orbital separation of $\sim$ 18.5 AU (Hatzes et al. 2003). 
Several other binary systems with separations of $\sim$ 20 AU appear to have
planetary companions, Gl 86 and HD 41004 A (Eggenberger et al. 2004).
However, of the binary or multiple systems with planets detected to 
date, most of the systems are considerably wider, with semimajor axes
ranging from $\sim$ 100 AU to $\sim$ 1000 AU or larger (Eggenberger et al. 
2004; Mayor et al. 2004; Mugrauer et al. 2004; Halbwachs et al. 2005). 
Three of the planet host stars are members of hierarchical triple systems, 
HD41004 A, HD 178911 B (Zucker et al. 2002), and 16 Cygni B, with the planet 
orbiting the single member of the triple system. Searches are underway for 
unknown binary companions to planet host stars, with the consequence
being that the number of planets found in binary or multiple
star systems is likely to increase as more companions are detected
(Patience et al. 2002; Mugrauer et al. 2004). Currently there are
at least 29 known binary or triple star systems with extrasolar planets
(M. Mugrauer 2004, private communication).

 Theoretical work on planet formation in binary systems has been
minimal because of the decades-long focus on understanding the
origin of the solar system. The discovery of extrasolar planets
in binary systems has now enlarged the theoretical realm to include
binary stars as well. Marzari \& Scholl (2000) and
Barbieri, Marzari, \& Scholl (2002) modeled the
formation of terrestrial planets in the $\alpha$ Centauri
binary star system, with a separation of $\sim$ 24 AU. They found
that while gravitational perturbations by the binary companion could
excite the eccentricities (and hence relative velocities) of planetesimals 
to values high enough to halt growth, the presence of gas drag 
introduces an orbital phasing that minimizes their relative
velocities and allow collisions to lead to growth rather than to
fragmentation, at least close ($\sim$ 2 AU) to one of the binary stars.
Using a symplectic integrator developed by Chambers et al. (2002), 
Quintana et al. (2002) modeled the final phase of growth of planetary
embryos into terrestrial planets in the $\alpha$ Centauri system,
finding that multiple terrestrial planets could form, provided
that the protoplanetary disk was inclined by no more than 60 degrees
to the plane of the binary system. Kortenkamp, Wetherill, \& Inaba 
(2001) found that a binary companion could serve as a 
source of orbital eccentricities leading to runaway growth
of planetary embryos into terrestrial planets, hastening the
formation process, as was also found by Quintana et al. (2002).

 Moriwaki \& Nakagawa (2004) extended the study of planetesimal
accretion to {\it circumbinary} protoplanetary disks, finding that for
a 1 AU binary separation and eccentricity $e$ = 0.1, planetesimals could only 
grow outside of 13 AU. Nelson (2003) studied the orbital evolution
of gas giant planets formed in circumbinary disks, finding that evolution
can lead to either ejection of the planet or to a stable orbit.
Marzari, Weidenschilling, Barbieri, \& Granata (2005) studied the orbital 
evolution of gas giant planets orbiting one of the stars in a binary 
system, finding that unstable initial conditions resulted in the hyperbolic
ejection of one or more planets, with the remaining planet being left
behind on an eccentric, shorter-period orbit.
 
 Th\'ebault et al. (2004) examined the formation of $\gamma$ Cephei's
gas giant planet in the core accretion scenario (Mizuno 1980), subject 
to the gravitational perturbations of the binary companion on a moderately
eccentric ($e$ = 0.36) orbit. They found that with a massive gaseous 
disk, needed to achieve orbital phasing, a 10 $M_\oplus$ core
could grow in $\sim$ 10 Myr, but that the core always ended up
at 1.5 AU, rather than out at the observed 2.1 AU.

 Nelson (2000) modeled the thermal and hydrodynamical evolution of
protoplanetary disks in an equal-mass binary system with a semimajor 
axis of 50 AU and $e$ = 0.3. The model was chosen to represent 
the L1551 IRS5 binary protostar system, where 0.05 $M_\odot$ disks orbit a 
pair of 0.5 $M_\odot$ protostars (Rodr\'iguez et al. 1998). Nelson (2000) 
found that following each periastron, the disks were heated by internal
shocks to such an extent that disk temperatures increased enough (to
$\sim$ 200 K at 10 AU) to not only prevent gas giant planet formation
by disk gravitational instability (Boss 1997), but also enough to
vaporize volatile solids and thereby prevent gas giant planet
formation by core accretion (Mizuno 1980). Nelson (2000) concluded that
``planet formation is unlikely in equal-mass binary systems
with $a \sim$ 50 AU.''

 Given the existence of several gas giant planets in binary
systems with separations of 20 AU or less, the negative results of 
Th\'ebault et al. (2004) and Nelson (2000) regarding the formation
of gas giant planets in binary systems clearly call for a re-examination
of this important question. The main thrust of this paper is
to present radiative hydrodynamical models of the disk instability 
mechanism for giant planet formation (Boss 1997, 2001, 2002a,b, 
2003) that add in the effects of a binary star companion.
Recent calculations with very high spatial resolution have shown
that the disk instability mechanism appears to become increasingly
vigorous as the continuum limit is approached (Boss 2005), and
furthermore that planets formed by this mechanism are relatively
immune to loss by orbital migration during a phase of gravitational
instability. We shall see that disk instability appears to be
capable of leading to the rapid formation of gas giant planets
in binary systems with a range of semimajor axes, provided that
the disk midplanes are cooled on an orbital time scale by vertical 
convection, as is indicated by similarly detailed models (Boss 2004a).
In fact, binary companions appear to be able to stimulate the
formation of self-gravitating protoplanets in otherwise
stable disks.

\section{Numerical Methods}

 The numerical calculations were performed with a finite volume 
hydrodynamics code that solves the three dimensional equations of 
hydrodynamics and the Poisson equation for the gravitational
potential. The same code has been used in many previous studies of
disk instability (Boss 2001, 2002a,b, 2003, 2004a, 2005) and has been shown
to be second-order-accurate in both space and time through convergence
testing (Boss \& Myhill 1992). The code has been tested on a variety of 
test cases (Boss \& Myhill 1992), including the nonisothermal test case for
protostellar collapse (Myhill \& Boss 1993). Bodenheimer et al. (2000)
found that the results obtained with this code agreed well with those
of an adaptive mesh refinement (AMR) code on isothermal collapse 
calculations.

 The equations are solved on a spherical coordinate grid with $N_r = 101$, 
$N_\theta = 23$ in $\pi/2 \ge \theta \ge 0$, and $N_\phi = 256$ or 512.
The radial grid is uniformly spaced with $\Delta r = 0.16$ AU
between 4 and 20 AU. The $\theta$ grid is compressed into the midplane to
ensure adequate vertical resolution ($\Delta \theta = 0.3^o$ at the midplane).
The $\phi$ grid is uniformly spaced, to prevent any bias in the azimuthal
direction. The central protostar wobbles in response to the growth of
nonaxisymmetry in the disk, thereby preserving the location of
the center of mass of the star and disk system. The number of terms in the
spherical harmonic expansion for the gravitational potential of the disk
is $N_{Ylm} = 32$ or 48. The Jeans length criterion (Boss 2002b)
is used to ensure that the clumps that form are not numerical
artifacts: even at the maximum clump densities, the numerical grid
spacings in all three coordinate directions remain less than 1/4 of
the local Jeans length.

 The boundary conditions are chosen at both 4 and 20 AU to absorb radial
velocity perturbations rather than to reflect mass and momentum back 
into the main grid (Boss 1998). Mass and linear or angular momentum entering 
the innermost shell of cells at 4 AU are added to the central protostar
and thereby removed from the hydrodynamical grid. No matter is allowed
to flow outward from the central cell back onto the main grid. Similarly,
mass and momentum that reaches the outermost shell of cells at 20 AU
piles up in this shell with zero radial velocity and is not allowed
to return to the main grid. The outermost gas does however continue to exert 
gravitational forces on the rest of the disk. 

 As in Boss (2001, 2002a,b, 2003, 2004a, 2005), the models treat radiative 
transfer in the diffusion approximation, which should be valid near the disk 
midplane and throughout most of the disk, because of the high vertical optical
depths. The divergence of the radiative flux term is set equal to zero 
in regions where the 
optical depth $\tau$ drops below 10, in order to ensure that the diffusion
approximation does not affect the solution in regions where it is not 
valid. As a result, it has not been found necessary to include a flux-limiter 
in the models (Boss 2001). The energy equation is solved explicitly in 
conservation law form, as are the four other hydrodynamic equations. 
Further details about the code may be found in Boss (2002b). 

\section{Artificial Viscosity}

 Artificial viscosity has not been used in the previous disk
instability models published by Boss (2001, 2002a,b, 2003, 2004a,
2005), but it has been included in a few models presented here
in order to explore its effects on clump formation. The implicit
artificial viscosity of this second-order accurate code, coupled
with small time steps (a result in part of the use of the spherical
coordinate system, rather than cylindrical coordinates), is sufficient 
to maintain stability of the code even in the presence of the strong
shocks driven by binary companions.

 Artificial viscosity can be used to help stabilize numerical schemes
and to provide microphysical heating within shocks. We use a tensor 
artificial viscosity (Tscharnuter \& Winkler 1979), which enters into the 
momentum equations as follows, where the other source terms on the right 
hand sides of these equations are suppressed for clarity,

$${\partial (\rho v_r) \over \partial t} + \nabla \cdot (\rho v_r {\vec v}) 
= ... - {1 \over r^3} {\partial (r^3 Q^r_r) \over \partial r},$$
$${\partial (\rho v_{\theta}) \over \partial t} + \nabla \cdot
(\rho v_{\theta} {\vec v}) = ... -
{1 \over r sin\theta } {\partial (sin\theta Q^\theta_\theta) \over
\partial \theta} + {Q^\phi_\phi cot\theta \over r},$$
$${\partial (\rho A) \over \partial t} + \nabla \cdot (\rho A {\vec v}) =
... - {\partial Q^\phi_\phi \over \partial \phi},$$

\noindent
where $\rho$ is the mass density, ${\vec v} = (v_r, v_{\theta}, v_{\phi})$
is the velocity, $A = r sin\theta v_{\phi}$ is the specific angular momentum, 
and the $Q^r_r$, $Q^\theta_\theta$, and $Q^\phi_\phi$ 
terms are the components of the artificial viscosity 
tensor. The artificial viscosity tensor is set equal to zero when
the divergence of the velocity field ($\nabla \cdot {\vec v}$) is positive
(i.e., in expanding regions), and when the divergence is negative, 
is defined to be

$$ Q^r_r =  l_r^2 \ \rho \ \nabla \cdot {\vec v} \
( {\partial v_r \over \partial r} - {1 \over 3}
\nabla \cdot {\vec v} ),$$
$$ Q^\theta_\theta = l_\theta^2 \ \rho \ \nabla \cdot {\vec v} \
( {1 \over r} {\partial v_\theta \over \partial \theta}
+ {v_r \over r} - {1 \over 3} \nabla \cdot {\vec v} ),$$
$$ Q^\phi_\phi = l_\phi^2 \ \rho \ \nabla \cdot {\vec v} \
( {1 \over r sin\theta} {\partial v_\phi \over \partial \phi}
+ {v_r \over r} + {v_\theta cot\theta \over r}
- {1 \over 3} \nabla \cdot {\vec v} ),$$

\noindent where $l_r^2 = max (C_r r^2, C_{\Delta r} \Delta r^2)$,
$l_\theta^2 = C_\theta (r \Delta \theta)^2$, and $l_\phi^2 = C_\phi
(r sin\theta\Delta \phi)^2$. $\Delta r$, $\Delta \theta$, and $\Delta \phi$ 
are the local grid spacings, $C_{\Delta r}$, $C_\theta$, and $C_\phi$
are free parameters usually set equal to 1, and $C_r$ is a free parameter
usually set equal to $10^{-4}$. The contribution to the right hand side 
of the specific internal energy equation is then

$$ E_Q= - Q^r_r \varepsilon^r_r 
- Q^\theta_\theta \varepsilon^\theta_\theta
- Q^\phi_\phi \varepsilon^\phi_\phi,$$

\noindent where

$$ \varepsilon^r_r = {\partial v_r \over \partial r},
\varepsilon^\theta_\theta = {1 \over r}
{\partial v_\theta \over \partial \theta} + {v_r \over r},
\varepsilon^\phi_\phi =  {1 \over r sin\theta}
{\partial v_\phi \over \partial \phi} + {v_r \over r} +
{v_\theta cot\theta \over r}.$$

\noindent $E_Q$ is constrained to be positive or zero,
reflecting the role of the artificial viscosity as a dissipative mechanism.
When artificial viscosity is to be used, the coefficient $C_\phi$ 
normally is set equal to zero in order to preserve the local
conservation of angular momentum. Only selected terms from the complete
tensor have been employed here. Terms involving coupling the
$r$ and $\theta$ components with the $\phi$ components have been
dropped (i.e., $Q^r_\phi$, $Q^\phi_r$, $Q^\theta_\phi$, and $Q^\phi_\theta$ 
are neglected), in order to conserve angular momentum locally in a 
consistent manner (see test cases in Boss \& Myhill 1992).

\section{Initial Conditions}

 The standard model consists of a $1 M_\odot$ central protostar 
surrounded by a disk with a mass of 0.091 $M_\odot$ between 4 and 20 AU. 
Disks with similar masses appear to be necessary to form gas giant 
planets by core accretion (e.g., Inaba, Wetherill, \& Ikoma 2003).
Most models also include the gravitational forces associated with
a $1 M_\odot$ binary star companion, as described below. Note that
the initial disk model does not include the gravitational forces
from the binary companion, so the evolution proceeds as if
the binary companion has just been formed, an unrealistic
but necessary assumption that is needed in order to make progress 
on this problem.

\subsection{Disk density}

 The initial protoplanetary disk structure is based on
the following approximate vertical density distribution (Boss 1993) for
an adiabatic, self-gravitating disk of arbitrary thickness in
near-Keplerian rotation about a point mass $M_s$

$$ \rho(R,Z)^{\gamma-1} = \rho_0(R)^{\gamma-1} $$
$$ - \biggl( { \gamma - 1 \over \gamma } \biggr) \biggl[
\biggl( { 2 \pi G \sigma(R) \over K } \biggr) Z +
{ G M_s \over K } \biggl( { 1 \over R } - { 1 \over (R^2 + Z^2)^{1/2} }
\biggr ) \biggr], $$

\noindent where $R$ and $Z$ are cylindrical coordinates, $\rho_0(R)$ is 
the midplane density, $G$ is the gravitational constant, and $\sigma(R)$ 
is the surface density. For setting up the initial model only, 
$K = 1.7 \times 10^{17}$ (cgs units) and $\gamma = 5/3$. The radial
variation of the midplane density is 

$$\rho_0(R) = \rho_{04} \biggl( {R_4 \over R} \biggr)^{3/2}, $$

\noindent where $\rho_{04} = 1.0 \times 10^{-10}$ g cm$^{-3}$ and
$R_4 = 4$ AU. 

\subsection{Disk temperatures}

 The initial temperature profile is based on two dimensional
radiative hydrodynamics calculations (Boss 1996) and is the
same as was used in previous models (Boss 2001, 2002a,b, 2004a). A
range of outer disk temperatures are investigated, with $T_o = 40$,
50, 60, 70, or 80 K (Table 1). As a result of the initial temperature and 
density profiles, the initial disks have $Q$ gravitational stability 
parameters whose minima range from $Q_{min} = 1.3$ for $T_o = 40$K.
to $Q_{min} = 1.9$ for $T_o = 80$K. [$Q$ is defined to be
$Q = c_s \Omega /(\pi G \sigma)$, where $c_s$ is the isothermal 
sound speed, $\Omega$ is the angular velocity, 
and $\sigma$ is the surface mass density of the disk.]
$T_o = 80$K is considerably higher than the temperatures at which the
solar system's comets are thought to have formed -- the experiments
of Notesco \& Bar-Nun (2005) imply that cometary nuclei agglomerated
from dust grains at $\sim$ 25 K, while observations of nuclear spin
temperatures of H$_2$O in three Oort Cloud comets suggest formation
temperatures of $\sim$ 20 to $\sim$ 45 K (Dello Russo et al. 2005). The 
Oort Cloud comets are thought to have formed between 5 and 40 AU, so they 
provide the ground truth for theoretical models of giant planet formation, 
at least in our planetary system. The outer disk temperatures of 
60, 70, and 80 K were then purposely chosen to be higher than 
expected for the solar nebula, in order to err on the conservative
side with regard to the outcome of a phase of disk instability.
Alternatively, models could be run with outer disk temperatures
closer to those inferred from comets, but with lower disk masses,
so that the initial values of $Q$ are again well above $\sim$ 1.5,
implying a relatively gravitationally stable initial disk. Models
starting with the same $Q$ values should evolve very similarly.

 In low optical depth regions, such as in the envelope infalling 
onto the disk, the temperature is assumed to be 50 K in the models, 
consistent with heating by radiation at distances of order 10 AU 
from a quiescent, solar-mass protostar (Chick \& Cassen 1997).
I.e., the disk surface is assumed to be immersed in a thermal bath
at a temperature of 50 K; the outer layers of the disk are thus
assumed to be able to radiate at whatever temperature is needed to
maintain this gas temperature. A more detailed calculation of
the thermal structure at the disk surface should be explored in
future models, as the surface temperature throttles disk cooling.
E.g., Chiang et al. (2001) calculated radiative, hydrostatic equilibrium
models of flared protoplanetary disks heated by radiation from their
central stars. Their two-layer disk models consisted of a disk surface
and a disk interior, with the optically thin disk surface being hotter 
than the disk interior, given the assumed heat source. At a distance of 
10 AU in their standard model, the disk surface temperature is 
$\sim$ 100 K and the interior temperature is $\sim$ 50 K. While the
gas and dust temperatures are roughly equal inside the disk, well above
the disk's photosphere the gas temperature can reach temperatures of
$\sim 10^4$ K (Kamp \& Dullemond 2004). Mechanical heating associated
with dynamical processes in the disk midplane may be the source of
the superheated atmospheres inferred for inner protoplanetary disks
(Glassgold, Najita, \& Igea 2004). At distances of 50 AU or more,
observations imply a vertical temperature gradient, with
midplane temperatures of $\sim$ 13-20 K underlying outer layers with 
temperatures of $\sim$ 30 K (Dartois, Dutrey, \& Guilloteau 2003).

\section{Binary Star Companion}

 The binary models include the gravitational accelerations from a binary 
star companion to the solar-mass star around that the disk orbits.
The models neglect any radiation coming from the second star in the
system.

\subsection{Tidal Potential}

 The tidal potential at a position $\vec r$ due to a binary star 
companion with mass $M_b$ located at $\vec r_b$ is given by

$$ \Phi_{tide}(\vec r) = - {G M_b \over | \vec r - \vec r_b |},$$

\noindent
where the binary star companion is represented
as a single point mass. The tidal 
potential may then be expressed in terms of an expansion in 
Legendre polynomials $P_l$ of order $l$ as

$$ \Phi_{tide}(\vec r) = - {G M_b \over r_b} 
\sum^{\infty}_{l = 0} \biggl( {r \over r_b} \biggr)^l P_l(cos S),$$

\noindent
where $S$ is the angle between $\vec r$ and $\vec r_b$. The $l = 1$
term in this expansion is responsible for the acceleration of
the primary star and its disk toward the binary companion,
an acceleration that is balanced by the centrifugal force
necessary for orbital motion of the primary star and its disk
around the center of mass of the entire system. Hence we take
as the tidal potential the following 

$$ \bar \Phi_{tide}(\vec r) = - {G M_b \over | \vec r - \vec r_b |}
 + {G M_b r \over r_b^2} cos S.$$

\noindent
The first non-trivial term in the tidal potential expansion will
then be the $l = 2$ term, which forces the disk into a 
prolate-ellipsoidal shape. When $\bar \Phi_{tide}(\vec r)$ is
added into the gravitational potential of the disk, obtained from
the solution of Poisson's equation, we have effectively included 
the tidal force of the orbiting binary companion (Boss 1981)
as well as the orbital motion of the star/disk around the center of
mass of the entire system. No other changes are needed for the
equations of motion (Mizuno \& Boss 1985). 

\subsection{Binary Star Orbit}

 We employ a nonrotating, noninertial reference frame for the 
models with a binary star companion, with the coordinate origin
fixed at the center of mass of the primary star and its disk.
Because of the way that the tidal force of the binary companion
has been included, in this reference frame the binary companion
appears to orbit around the coordinate origin of the disk
whose evolution is being calculated (Mizuno \& Boss 1985). A
similar approach was used by Larwood et al. (1996) in their
models of accretion disks being warped by binary companions.
In the present models, the binary star is assumed 
to lie in the same plane as the disk, so that no warps are
created, and the disk retains its symmetry above and below
its midplane.

 The binary star companion is assumed to be on an orbit
with eccentricity $e_b$ and semimajor axis $a_b$ (Table 1).
$\phi_{bi}$ defines the initial position angle of the binary in
its eccentric orbit, with $\phi_{bi} = 0$ corresponding to
starting at periastron, and $\phi_{bi} = \pi$ to apoastron.
$\phi_{b}(t)$ denotes the position angle of the companion
as it moves along its orbit (i.e., the true anomaly, $f$).
For Keplerian orbits, $\phi_{b}(t)$ is calculated by

$$ \phi_{b}(t) = \phi_{bi} + \int^t_0 \biggl[ {J_b \over r_b^2(t)} 
\biggr] dt, $$ 

\noindent
where the angular momentum per unit mass $J_b$, a constant, is equal to
  
$$J_b = \Omega_b a_b^2 (1 - e_b^2)^{1/2}.$$

\noindent
$\Omega_b$, the mean motion, is equal to $\Omega_b = 2 \pi/P_b$, where
$P_b$ is the orbital period of the binary. The mean motion is
also equal to

$$\Omega_b = \biggl( { G (M_s + M_b) \over a_b^3 } \biggr)^{1/2}, $$

\noindent
where $M_s$ is the mass of the star with the disk. The binary separation
$r_b(t)$ is determined from the time evolution of $\phi_b(t)$ through

$$r_b(t) = a_b {(1 - e_b^2) \over (1 + e_b cos \phi_b(t))}. $$

\noindent
  
In these models, an equal mass binary system is assumed, i.e.,
$M_s = M_b = 1 M_\odot$. The only free parameters then are
$a_b$, $e_b$, and $\phi_{bi}$, as noted in Table 1.
Models with $\phi_{bi} = 0$ start at periastron, so that
$r_b(t = 0) = a_b (1 - e_b)$, whereas models 
with $\phi_{bi} = \pi$ start at apoastron, so that
$r_b(t = 0) = a_b (1 + e_b)$. In order to avoid abrupt initial
changes in the disk when starting a model, the tidal forces
begin at zero strength and increase linearly with time over the first
30 yrs of evolution, when their full strength is reached and
maintained thereafter.
 
\section{Results}

 Table 1 summarizes the disk models with and without binary companions.
The latter models are presented here in order to be able to separate
out the effects of including the binary companions from what the
disks would do in the absence of external forces.

\subsection{Models without Binary Companions}

 We begin with several disk instability models that are identical
to those previously published by Boss (2002), except for starting
with higher initial outer disk temperatures ($T_o$). Boss (2002)
presented results for models with initial $T_o =$ 40K and 50K (as in
models eb and ab), leading to initial Toomre (1964) $Q$ stability 
values of $Q_{min}$ = 1.3 and 1.5, respectively. In these initially
marginally gravitationally unstable disks, strongly nonaxisymmtric 
structures begin to form within a few hundred years of evolution,
equal to about 10 orbital periods at an orbital radius of $\sim$ 10 AU
where clumps first appear in an unperturbed disk of this type. Given 
that the orbital period of the $a_b = 50$ AU binary system is 250 yrs,
it is clear that the unperturbed disks with initial $Q_{min}$ = 1.3 and 1.5
can be expected to develop nonaxisymmetry on the same time
scale as the binary perturbations. Hence models were studied
with higher initial temperatures in order to try to see what
would happen in a disk that might not do much on its own prior
to being excited by the binary perturber. Models f, g, and h thus
began with $T_o =$ 60K, 70K, and 80K, respectively,
leading to an initial $Q_{min}$ = 1.6, 1.8, and 1.9. These
models are more gravitationally stable initially than models
with $T_o =$ 40K and 50K, and hence should also represent a
protoplanetary disk that has not yet evolved into a state of
marginal gravitational instability.

Figure 1 shows the initial radial distribution of the surface
density in model f with $T_o =$ 60K, compared to the critical
surface density needed to make the disk have a Toomre $Q = 1$
at that radius, i.e., in order to be strongly gravitationally
unstable initially. Because the initial temperature profile
rises high above $T_o$ inside $\sim$ 8 AU, this critical surface
density rises sharply as well. Hence the innermost regions
are expected to remain gravitationally stable. Figure 2 shows
the surface density for model f after 87.1 yrs of evolution.
The presence of axisymmetric rings and growing spiral arms
can be inferred from the ripples in the surface density.
In addition, it is clear that the region inside $\sim$ 6 AU
has already been significantly modified from the initial
profile, with mass having been transported inward onto the central
protostar as well as outward to the growing ring centered around 6 AU.
Figures 3 and 4 show the further evolution of model f after 160 yrs and
233 yrs, respectively, as the innermost region is severely
depleted of gas and dense ring-like features grow between 8 AU
and 10 AU. The high average surface density at the 4 AU inner boundary 
is produced by a few dense cells where disk mass is flowing onto
the central protostar and exiting the hydro grid. These figures
show that the disk evolves to form rings that become increasingly
closer to Toomre $Q = 1$ instability [in fact, the Toomre (1964) 
criterion explicitly refers to ring formation as a predecessor
to the development of nonaxisymmetry.]

 This trend is further displayed in Figures 5 and 6, which show the
evolution of the Toomre $Q$ parameter for model h ($T_o = 80$ K).
Starting from a disk with a minimum Toomre $Q$ value of 1.9,
considerably stabler than model f with 1.6, Figure 6 shows that
after 245 yrs of evolution the disk has formed rings around 10 AU
where $Q$ drops to $\sim 1.5$, sufficient for marginal gravitational
instability. The inner regions become even more stable ($Q > 2$)
as a result of their higher temperatures and depleted gas surface
density. Figures 1-6 make it clear why clumps tend to form
preferentially around 10 AU in these models, as interior to that
distance is where midplane temperatures rise to higher values at 
smaller radii and where the disk surface density is depleted by 
accretion onto the central protostar. Beyond 10 AU the instability
proceeds somewhat slower because of the longer orbital time
periods.

 Figures 7 and 8 show the formation of clumps in models f and h
at times of 233 yrs and 245 yrs, respectively. While the clumps
are not necessarily self-gravitating at this phase of evolution,
it is clear that these disks are trying to form clumps in spite
of their relatively high initial outer disk temperatures, higher
in fact than appears to be appropriate for the solar nebula
based on cometary speciation (e.g., Dello Russo et al. 2005). 

 Evidently even low amplitude nonaxisymmetry can transfer
mass and angular momentum over times of order 10 orbital
periods sufficient to approach a more robust phase of gravitational
instability. Models f, g, and h suggest that the natural 
evolution of gravitationally stable disks is toward marginal
gravitational instability and then on to clump formation,
even in the absence of triggering effects such as binary
companions or secular cooling. Note that in all of these models the
outer disk temperature is not allowed to drop below the initial 
value of $T_o$, in an attempt to err on the side of being 
conservative with respect to thermal decompression and cooling.

\subsection{Models with Binary Companions}

 Table 1 lists final times $t_f$ of the models with binary
star companions on orbits with eccentricity $e_b$ and semimajor 
axis $a_b$. For models with $a_b$ = 50 AU, the binary orbital
period is 250 yrs, while for $a_b$ = 100 AU, it is 707 yrs
(note that the binary companion is also a solar-mass star).
The evolution of the models following the first periastron
passage of the binary companion should also be relevant for the
problem of a disk around a single star that undergoes a very
close encounter with another star in the star-forming cluster.

 Figures 9-12 show the time evolution of model gbca, where
the binary companion has $a_b$ = 50 AU and $e_b$ = 0.5. The disk
models start out essentially axisymmetric with only a low
level ($\sim$ 1\%) of noise. After 83.8 yrs of evolution,
the disk has become slightly nonaxisymmetric (Figure 10),
primarily as a result of its own evolution (see previous
section). However, by 139 yrs (Figure 11), the binary
companion has completed just over half of an orbital
period and has passed periastron at a distance of 25 AU
from the center of the disk, severely distorting the outer
regions of the disk (note that the density concentrations
at 20 AU are an artifact of the disk boundary conditions
at 20 AU, where disk material is allowed to enter the
outermost shell of cells but cannot flow farther away
as it would in a calculation with a more distant outer
boundary.) The binary companion is located at this time
at about 2 o'clock. Periastron was at 3 o'clock, and the
binary companion orbits in a counter-clockwise sense, in the same 
direction that the disk gas orbits, consistent with formation of the
entire system from a single rotating, dense cloud core.
 
 While the structures in the outermost disk are strongly influenced
by the outer boundary conditions, the innermost arcs are not.
The tidal forces of the binary companion have forced the disk
into a prolate shape that is beginning to wind-up in the
inner regions because of Keplerian rotation (Figure 11). By
the time of Figure 12, at 191 yrs, the binary companion is
approaching apoastron, but the tidally perturbed disk is
still forming spiral arms and clumps, as well as strong
shock fronts in its innermost regions. Clearly the presence
of a binary companion with these orbital parameters has had
a major effect on the evolution of this initially gravitationally
stable disk, inducing the formation of clumps after the first
periastron. This fact makes it clear that starting this model
with an axisymmetric disk is not correct -- in a real disk orbiting 
a protostar in a binary system of this type, the outer disk is never
axisymmetric. Axisymmetric initial models are a theoretical
convenience that allows one to jump into the system in an
approximate manner and to follow the subsequent evolution.

 Figure 13 shows the midplane density contours for model ab
after 159 yrs. The disk has already been tidally perturbed because
this model began with the binary companion at periastron, though
the tidal forces were turned on over a time period of 30 yrs.
A well-defined clump is evident at 6 o'clock in Figure 13,
containing $\sim 1.5 M_{Jup}$ of gas and dust. This clump is
sufficiently massive to be gravitationally bound, as the
Jeans mass at the mean density ($7.2 \times 10^{-10}$ g cm$^{-3}$)
and temperature (161 K) of this clump is only 1.4 $M_{Jup}$.
The spherically-averaged radius of the clump is 0.66 AU, 
only slightly larger than the tidal stability radius of 0.64 AU, 
implying marginal tidal stability. The clump is moving on
an orbit with $a = 8.2$ AU and $e = 0.094$ at this time.

 Figure 14 demonstrates that the clump in Figure 13 is properly
resolved with respect to the Jeans length criterion, which 
dips to close to the grid resolution at the location of the
clump's density maximum, seen in Figure 15. Figure 16 shows
how the temperatures within the clump have risen considerably
over the initial temperatures as a result of compressional
heating -- the maximum temperature in the clump exceeds 300 K,
compared to a mean temperature of 161 K. Figure 17 shows the
temperature distribution thoughout the midplane of model ab
after 159 yrs, showing the effects of heating throughout the
disk. The disk is vertically unstable to convection according
to the Schwarzschild criterion at the location of the dense
clump seen in Figure 13, as well as at a number of other
radii near the midplane in model ab. Convective cooling appears
to be important for transporting thermal energy from the disk
midplane to the disk atmosphere, where it can be radiated
away, allowing a disk instability to produce dense clumps centered
on the midplane (Boss 2004a).
  
 The effect of the binary eccentricity on the models can be
seen by comparing Figures 18 and 19. Models hbcae (Figure 18)
and hbca (Figure 19) are identical except that the binary
eccentricity is 0.25 for the former and 0.5 for the latter
($a_b = 50$ AU for both models). As a consequence, periastron
occurs at a radius of 37.5 AU for model hbcae and at 25 AU
for model hbca, leading to considerably stronger tidal
forces in the latter model. Figures 18 and 18 demonstrate
this point after one binary orbital period has elapsed: while
both disks have formed strong spiral arms and clumps,
model hbca is clearly more strongly distorted and has
developed higher densities along the outer boundary of the disk.
The clump at 10 o'clock in Figure 18 has a mass of
4.7 $M_{Jup}$, sufficiently high to be strongly self-gravitating,
whereas the clump at  2 o'clock in Figure 19 is not
quite self-gravitating with a mass of 0.68 $M_{Jup}$.
This suggests that while binary perturbers can stimulate
clump formation, too strong of a perturbation can make it
harder for the clumps to survive to become true protoplanets.
However, even in model hbca, other clumps form later
in the evolution that are dense enough and massive enough
to be self-gravitating.

 The effect of the binary semimajor axis on the models can be
seen by comparing Figures 20 and 21. Models gba (Figure 20)
and gbca (Figure 21) are identical except that the semimajor
axis is 100 AU for the former and 50 AU for the latter
($e_b = 0.5$ AU for both models). As a consequence, periastron
occurs at a radius of 50 AU for model gba and at 25 AU
for model gbca, again leading to considerably stronger tidal
forces in the latter model. Figures 20 and 21 demonstrate
the effects of these different semimajor axes, shortly after
one binary orbital period has elapsed, in order to compare
these models at an equivalent time with respect to the
effects of the tidal forces. While model gba has evolved and
formed spiral arms, dense clumps have not formed at this time. 
The evolution is closer to that of the disk models without
binary companions -- evidently tidal forces from binary companions 
at distances of $\sim$ 50 AU or greater have relatively little
effect on the disk inside 20 AU. In model gbca (Figure 21),
on the other hand, the binary's periastron of 25 AU has had
a major effect on the disk, and has induced the formation of
a dense clump at 9 o'clock with a mass of 1.7 $M_{Jup}$.  
Strong spiral arms are also evident throughout the disk.

 In order to ascertain the effects of the numerical resolution, 
model gbca was continued from the time shown in Figure 21 as 
model gbcah with double the number of azimuthal grid
points (i.e., $N_\phi = 512$ instead of $N_\phi = 256$),
and more terms in the gravitational potential solution
(i.e., $N_{Ylm} = 48$ instead of $N_{Ylm} = 32$). Model gbcah
is shown in Figure 22 after another 28 yrs of evolution beyond
the point shown in Figure 21, i.e., roughly another orbital
period at $\sim$ 10 AU. A self-gravitating clump orbits at 10 AU
(seen at 8 o'clock) with a mass of 1.2 $M_{Jup}$, well above
the relevant Jeans mass of 0.72 $M_{Jup}$, with a radius (0.76 AU) 
comparable to the tidal stability radius (0.75 AU). Clump
formation and survival is enhanced as the spatial resolution
is increased in the critical azimuthal direction (Boss 2000, 2005).

\subsection{Models with Varied Thermodynamical Stability Handling}

Two approaches have been used in these models and in the previous disk
instability models by Boss (2001, 2002a,b, 2003, 2005) for stability of 
the radiative transfer solution, given the use of an explicit
time differencing scheme for the solution of the energy equation
in the diffusion approximation. First, taking smaller time steps
(i.e., smaller fractions of the Courant time) is often sufficient
to maintain stability of the thermodynamical solution. The calculations
typically begin with a time step that is 50\% of the minimum
Courant time on the grid. For some models, this fraction is reduced
to maintain stability, to values as small as 1\%, though typically
the fraction remains no smaller than 5\% or 10\%. While sufficient
to maintain stability, clearly this approach slows the calculation
proportionately. Hence it has been found useful to use a numerical
artifice to try to maintain stability of the numerical solution of
the energy equation in the low density regions where it tends
to break down. The artifice is simple: when the density inside
the disk drops below a specified critical value, $\rho_{crit}$,
then the temperature in that cell is forced back to its initial
temperature at the beginning of the evolution. This artifice is
justified to the extent that such regions are low in density
because they are undergoing decompression, and hence should also
be undergoing decompressional cooling. Setting the temperatures
of such regions to a value no lower than their initial temperature
is then a relatively conservative approach.

While the question of the handling of $\rho_{crit}$ may seem to be
largely a technical point, given the sensitivity of
the outcomes of disk instability calculations to the heating and cooling
processes in the disk, it is important to examine any technical
details that might have an unintended effect on the results.

All the models began with $\rho_{crit} = 10^{-13}$ g cm$^{-3}$, compared
to the initial midplane density of $10^{-10}$ g cm$^{-3}$ at 4 AU.
In order to maintain stability with a reasonably-sized time step,
however, in some models $\rho_{crit}$ is increased to values of
$3 \times 10^{-12}$ g cm$^{-3}$ or $10^{-11}$ g cm$^{-3}$. 
With these values, even moderately low density regions of the disk
are effectively forced to behave isothermally. With this in mind, all
the models were searched for evidence that the highest value
of $\rho_{crit}$ used had a significant effect on the outcome
of the evolution. The primary criterion employed was looking
for the maximum density produced in the disk midplane around
5 AU to 10 AU, where the dense clumps form. It was found that
the maximum density reached was typically the same
($\sim 2 \times 10^{-9}$ g cm$^{-3}$) independent of whether
$\rho_{crit}$ stayed at a value of $10^{-13}$ g cm$^{-3}$ 
throughout the calculation, or had to be increased at some point
to $3 \times 10^{-12}$ g cm$^{-3}$ or $10^{-11}$ g cm$^{-3}$
to maintain a stable solution. This results suggests that
the $\rho_{crit}$ artifice is not a major 
determinant of the outcome.

\subsection{Models with Artificial Viscosity}

 Hydrodynamical calculations where artificial viscosity is employed 
generally have not found robust clump formation in either fully three 
dimensional (Pickett et al. 2000) or in thin disk models (Nelson 2000). 
Here we show that when artificial viscosity is included in three dimensional 
disk models with radiative and convective cooling, the tendency to form 
clumps is reduced somewhat, but not eliminated, unless the artificial 
viscosity is increased by a factor of order ten. 

 These models have the standard spatial resolution (Boss 2002b) of 100 
radial grid points distributed uniformly between 4 and 20 AU, 256 azimuthal 
grid points, 22 theta grid points in a hemisphere (effectively over a million 
grid points), and include terms up to $l,m = 32$ in the spherical harmonic
solution for the gravitational potential. The models begin after 322 years 
of inviscid evolution of a disk with an initial mass of 0.091 $M_\odot$ 
(Boss 2002b), an outer disk temperature of 40 K, and a 
minimum Toomre $Q = 1.3$.

 Figures 23 through 26 show the results for four models that are identical 
except for their treatment of artificial viscosity. It can be seen that
in the models with the standard artificial viscosity 
(Figure 24: $C_{\Delta r} = C_\theta = 1$, $C_\phi = 0$, $C_r = 10^{-4}$; 
Figure 25: same as Figure 24, but with $C_\phi = 1$), 
clump formation occurs in a similar manner as in the model without 
artificial viscosity (Figure 23, as in Boss 2002b). However, when
the artificial viscosity is increased by a factor of 10 (Figure 26), clump
formation is significantly inhibited because of the heating associated
with the assumed dissipation. These models support the suggestion that
microphysical shock heating can be important for clump formation (Pickett 
et al. 2000), though with the standard amount of artificial viscosity, 
the effects are relatively minor in these models. Calibrating the
proper amount of artificial viscosity that would be needed to properly
represent the correct level of microphysical (sub-grid) shock heating
remains as a challenge, but it is clear that large amounts of
artificial viscosity can suppress clump formation.

\section{Discussion}

\subsection{HD 188753 triple star system}

 Recently Konacki (2005) has claimed the discovery of a hot Jupiter 
in orbit around a 1.06 $M_\odot$ star that is a member of 
the hierarchical triple star system HD 188753.
The average distance between the primary star and the binary secondary
is 12.3 AU, with the secondary being on an orbit
with $e = 0.5$ and having a total mass of 1.63 $M_\odot$. This means
that at periastron, the secondary passes within $\sim 6$ AU of
the primary, rendering orbits outside of $\sim 1.5$ AU unstable. 
Hot Jupiters are thought to form at several AU from solar-mass
stars and then to migrate inward to short-period orbits by gravitational
interactions with the gaseous disk. However, the protoplanetary disk 
around the primary star in HD 188753 would be restricted in extent 
to $\sim 1.5$ AU and so could not extend out to regions
cool enough for icy grains to contribute to assembling the solid core
required for the core accretion mechanism or cool enough
for a disk instability to occur. Given the difficulty of 
forming gas giant planets {\it in situ} on short period
orbits by either core accretion (Bodenheimer, Hubickyj, \& Lissauer 2000)
or disk instability (Boss 1997), the presence of the planet in
HD 188753 is thus puzzling, given the current orbital configuration,
if the discovery can be confirmed.

 However, the fact that HD 188753 is a triple system offers a
possible solution. Hierarchical triples can form by the orbital evolution 
of an initially equally-spaced multiple protostar system (e.g., Boss 2000).
This evolution proceeds over a period of $\sim 100$ orbital crossing
periods. For a multiple protostar system with an initial separation
of $\sim 100$ AU, the initial orbital period would be $\sim 10^3$ yrs,
so that the initial equally-spaced multiple protostar system would be
expected to undergo a series of close encounters and ejections
leading to the final, stable, hierarchical triple system within a
time period of $\sim 10^5$ yrs. If a gas giant planet could form
within the protoplanetary disk of one of the protostars within
$\sim 10^5$ yrs, it might then survive the subsequent orbital
evolution as a hot Jupiter. Rapid formation is required, suggesting
that a disk instability might be needed to explain HD 188753's
putative hot Jupiter. 

\subsection{Previous calculations}

 Contrary to the results of Nelson (2000), these models suggest
that tidal forces from binary companions need not prevent the
formation of giant planets, by either the disk instability or
core accretion mechanisms. The key difference is in the midplane
temperatures reached after periastrons, with the Nelson (2000)
models reaching temperatures high enough to sublimate icy dust
grains at $\sim 10$ AU and to prevent a robust disk instability
inside this radius. Here we try to understand why the present
results differ from those of Nelson (2000).

There are several important similarities and differences between the two 
sets of calculations. Nelson (2000) used 60,000 SPH (smoothed particle
hydrodynamics) particles in each disk, compared
to effectively over $10^6$ grid points in the present models 
with $N_\phi = 256$, though because Nelson's calculations were 
restricted to two dimensional (thin) disks, the spatial resolution 
was similar to that in the midplane of the present models with 
$N_\phi = 512$. Nelson (2000) assumed a thin disk with
an adiabatic vertical temperature gradient, which assumes that
vertical convection is able to keep the vertical temperature gradient
at the adiabatic level. This results in the maximum possible temperature
difference between the disk surface (excluding the disk photosphere)
and the midplane, because if radiative transport were efficient, the 
vertical temperature gradient would not be as steep. The present
models start out vertically isothermal, but then develop vertical 
convective motions in regions where the vertical temperature gradient
exceeds the adiabatic value (i.e., the Schwarzschild criterion for
convection is met; Boss 2004a). Nelson (2000) also used disk surface 
temperatures (100 K) greater than those assumed in the present models 
(50 K), leading to higher midplane temperatures, though the higher
surface temperatures should lead to a higher rate of radiative cooling.

 Perhaps the most likely source of the discrepancy is the amount
of artificial viscosity assumed in the two sets of models.  
Artificial viscosity equivalent to an effective alpha viscosity
with $\alpha = $ 0.002 to 0.005 was intentionally included in the 
Nelson (2000) models in an effort to include the effects of shocks 
and sub-grid turbulence. In the present models, artificial viscosity is
not used, and the degree of implicit numerical viscosity appears
to be at a level equivalent to $\alpha \sim 10^{-4}$ (Boss 2004b),
a factor of 20 to 50 times lower than that in Nelson (2000).
As we have seen in Figures 23-26, a high level
of artificial viscosity can heat the disk sufficiently to suppress the 
formation of clumps, though Figures 24 and 25 show that with a standard 
amount of artificial viscosity, clumps can still form. The artificial 
viscosity employed in SPH codes can lead to a ``large and unphysical
shear dissipation as a side effect in disk simulations'' (Nelson et al.
2000), though Nelson et al. (2000) and Nelson (2000) used a
formulation that was intended to minimize artificial viscous 
dissipation. Nevertheless, the intentional use of a relatively
large amount of artificial viscosity (in order to attempt to
duplicate spectral energy distributions for observed disks)
is likely to be the main source of the discrepancy between the
models. This artificial viscous heating appears to be related to 
the difference in cooling times in the two sets of models, as
the cooling time is critical for clump formation and survival.
Relatively short cooling times are obtained in the present models 
($\sim 1$ to 2 orbital periods, Boss 2004a), compared to the effective 
cooling time obtained in Nelson (2000) of $\sim 5$ to $\sim 15$ 
orbital periods for distances from 10 AU to 5 AU, respectively 
(Nelson 2005, private communication).

 One could reasonably ask whether the present models are able
to handle strong shocks properly in the absence of artificial
viscosity, as that is how these models have been run, with the
exception of the models shown in Figures 24-26. In order to
test this possibility, one dimensional shock tests performed with the
same hydrodynamic scheme as used in the present models and first 
presented by Boss \& Myhill (1992) were repeated with and without
artificial viscosity. The shock test relies on the analytic
solution for the Burgers equation presented by Harten \& Zwas
(1972). Using the same numerical code and numerical parameters
as presented in Figure 7 of Boss \& Myhill (1992), Figures 27 and
28 depict the results with the standard amount of artificial
viscosity ($C_Q = 1$) and with zero artificial viscosity, respectively. 
It can be seen that in both cases, the numerical solution
does an excellent job of reproducing the analytical solution,
including the shock front location. Figure 28 shows that
in the complete absence of artificial viscosity, there is a
similar degree of overshoot/undershoot immediately downstream of the 
shock front as in Figure 27 with non-zero artificial viscosity (in 
both cases, the overshoot/undershoot is minimal compared to that of 
several other differencing schemes; see Figure 7 of Boss \& Myhill 1992).
These results suggest that the present models, even with zero
artificial viscosity, are able to handle strong shocks about as
well as if the standard amount of artificial viscosity were
being employed. It is thus likely that with the standard amount
of artificial viscosity, the effective $\alpha$ of the models
is similar to that caused by the implicit numerical 
viscosity ($\alpha \sim 10^{-4}$; Boss 2004b). In that case,
the Nelson (2000) models effectively include viscous dissipation
at a rate roughly 20 to 50 times higher than the present models,
which appears to be sufficient to explain the suppression of clump 
formation in the Nelson (2000) models, based on the results presented 
in Figures 23-26.

 There may be a related discrepancy between these models 
and those of Nelson (2000). Nelson (2000) found that the long-wavelength
flux densities from his disk models were below those measured for the
L1551 IRS5 binary disk system upon which his models were based,
implying effective temperatures for the disk surface that were too low. 
However, Boss \& Yorke (1993, 1996) found that they were able
to match the spectral energy distributions of the T Tauri system with
the same axisymmetric disk models that form the basis for the three dimensional
disk models used in the present models. It is unclear at present what
this means, but suffice it to say that a higher effective temperature
at the disk surface should increase radiative losses from the disk surface
and thereby reduce the overall disk cooling time, though perhaps
at the expense of higher midplane temperatures. 

\section{Conclusions}

 These models have shown that initially stable protoplanetary
disks can evolve over time periods of $\sim 10^3$ yrs to become
marginally gravitationally unstable and then begin to form clumps.
When these stable disks are perturbed by strong tidal forces
(i.e., periastrons less than $\sim$ 50 AU), spiral arms form
soon after peristron and typically evolve into self-gravitating,
dense clumps capable of forming gas giant planets. Periastrons
of $\sim$ 50 AU and larger lead to little effect on the
evolution of these disks, which are limited in extent to 20 AU.
Disk cooling processes such as convection appear to remain
effective enough to permit self-gravitating clumps to form,
even in the presence of the strong tidal forcing. As a result,
outer disk temperatures do not become high enough in general
for icy dust grains to be sublimated, meaning that giant planet
formation by core accretion would continue to be aided by
the enhanced surface density of solids associated with the ice
condensation boundary in the disk, even in binary star systems.
Given the tendency for these disks to form self-gravitating clumps
by disk instability on a time scale of $\sim 10^3$ yrs or less,
these models suggest that giant planets should be able to
form in binary systems with periastrons as small as 25 AU,
by either core accretion or disk instability. This general
conclusion seems to be consistent with the growing observational
evidence for giant planets in binary star systems.

 Because of the nature of a spherical coordinate grid, 
where $\Delta x_\phi = r sin \theta \Delta \phi$ increases linearly
with radius, the present models often fail to properly resolve any 
clumps that try to form near the edge of the edge. An improved
treatment of disks being strongly perturbed by binary companions
would require the use of adaptive mesh refinement (AMR) code or
some other technique for better resolving clumps at large radii.

 I thank Andy Nelson for details about the cooling times in his 
calculations, the referee for extremely helpful comments and 
questions about artificial viscosity and viscous dissipation,
and Sandy Keiser for her continued expert assistance with the 
Carnegie Alpha Cluster. This research was supported in part by the NASA 
Planetary Geology and Geophysics Program under grant NNG05GH30G, 
by the NASA Origins of Solar Systems Program under grant NNG05GI10G,
and by the NASA Astrobiology Institute under grant NCC2-1056.
Calculations were performed on the Carnegie Alpha Cluster, the purchase
of which was supported in part by NSF MRI grant AST-9976645.

\vfill\eject

\begin{figure}
\vspace{-2.0in}
\caption{A copy of the paper that includes the figures may be
downloaded from http://www.dtm.ciw.edu/boss/ftp/bin/.}
\end{figure}

\clearpage
\begin{deluxetable}{ccccccccc}
\tablecaption{Summary of disk instability models with and without
binary companions. \label{tbl-1}}

\tablehead{\colhead{model} &
\colhead{\quad $T_o$ (K) } &
\colhead{$min(Q_i)$ } &
\colhead{$a_b$ (AU) } &
\colhead{\quad $e_b$ } &
\colhead{\quad $\phi_{bi}$ } &
\colhead{\quad $N_\phi$ } &
\colhead{\quad $N_{Ylm}$ } &
\colhead{$t_{f}$ (yrs) } }

\startdata

eb   & 40K & 1.3 & 100 & 0.5  &     0  &  256 & 32 &   610 \\

\hline

ab   & 50K & 1.5 & 100 & 0.5  &     0  &  256 & 32 &   616 \\    

\hline

f    & 60K & 1.6 &  -- &  --  &     -- &  256 & 32 &  1139 \\ 
fb   & 60K & 1.6 & 100 & 0.5  &     0  &  256 & 32 &   938 \\  
fbc  & 60K & 1.6 &  50 & 0.5  &     0  &  256 & 32 &   831 \\
fbca & 60K & 1.6 &  50 & 0.5  &  $\pi$ &  256 & 32 &  1126 \\   

\hline

g    & 70K & 1.8 &  -- &  --  &     -- &  256 & 32 &   891 \\
gbae & 70K & 1.8 & 100 & 0.25 &  $\pi$ &  256 & 32 &   616 \\ 
gb   & 70K & 1.8 & 100 & 0.5  &     0  &  256 & 32 &  1018 \\
gba  & 70K & 1.8 & 100 & 0.5  &  $\pi$ &  256 & 32 &  1481 \\ 
gbcae& 70K & 1.8 &  50 & 0.25 &  $\pi$ &  256 & 32 &  1059 \\
gbca & 70K & 1.8 &  50 & 0.5  &  $\pi$ &  256 & 32 &  1146 \\
gbcah& 70K & 1.8 &  50 & 0.5  &  $\pi$ &  512 & 48 &   630 \\
gbc  & 70K & 1.8 &  50 & 0.5  &     0  &  256 & 32 &   744 \\
gbch & 70K & 1.8 &  50 & 0.5  &     0  &  512 & 48 &   429 \\

\hline

h    & 80K & 1.9 &  -- &  --  &    --  &  256 & 32 &  1970 \\
hbae & 80K & 1.9 & 100 & 0.25 & $\pi$  &  256 & 32 &  1903 \\ 
hba  & 80K & 1.9 & 100 & 0.5  & $\pi$  &  256 & 32 &  1246 \\
hbcae& 80K & 1.9 &  50 & 0.25 & $\pi$  &  256 & 32 &  1910 \\
hbca & 80K & 1.9 &  50 & 0.5  & $\pi$  &  256 & 32 &   657 \\

\enddata
\end{deluxetable}
\clearpage

\suppressfloats

\end{document}